# Finding People's Professions and Nationalities Using Distant Supervision

The FMI@SU "goosefoot" team at the WSDM Cup 2017 Triple Scoring Task

Valentin Zmiycharov
FMI, Sofia University "St. Kliment Ohridski"
Sofia, Bulgaria
valentin.zmiycharov@gmail.com

Dimitar Alexandrov
FMI, Sofia University "St. Kliment Ohridski"
Sofia, Bulgaria
dimityr.alexandrov@gmail.com

Preslav Nakov
Qatar Computing Research Institute, HBKU
Doha, Qatar
pnakov@qf.org.qa

Ivan Koychev
FMI, Sofia University "St. Kliment Ohridski"
Sofia, Bulgaria
koychev@fmi.uni-sofia.bg

Yasen Kiprov
FMI, Sofia University "St. Kliment Ohridski"
Sofia, Bulgaria
yasen.kiprov@gmail.com

## ABSTRACT

We describe the system that our FMI@SU student's team built for participating in the Triple Scoring task at the WSDM Cup 2017. Given a triple from a "type-like" relation, *profession* or *nationality*, the goal is to produce a score, on a scale from 0 to 7, that measures the relevance of the statement expressed by the triple: e.g., how well does the *profession* of an *Actor* fit for *Quentin Tarantino*? We propose a distant supervision approach using information crawled from Wikipedia, DeletionPedia, and DBpedia, together with task-specific word embeddings, TF-IDF weights, and role occurrence order, which we combine in a linear regression model. The official evaluation ranked our submission 1st on Kendall's Tau, 7th on Average score difference, and 9th on Accuracy, out of 21 participating teams.

## 1. INTRODUCTION

Given a triple from a "type-like" relation, the shared task[1] asks participants to compute a score, on a discrete scale from 0 to 7, that measures the relevance of the statement expressed by the triple compared to other triples from the same relation. The task focuses on two such relations: *profession* and *nationality*. The *type-like* relation is defined as follows:

(<Entity Name>, <Relation Type>, <Relevance score>)

where *Entity Name* corresponds to the name of the person, *Relation Type* can be *profession* or *nationality*, and *Relevance score* measures how close the person is to the relation. Here are two examples for the nationality relation:

    *Romano Scavolini Croatia    1*
    *Romano Scavolini Italy    7*

The organizers provided 33,159,353 sentences, including annotations of 385,426 persons. Only people from the already provided list were used in the test sets. The participating systems were evaluated against two test sets: one for professions and one for nationalities.

The task puts no limitations on the kind or amount of additional external training data that could be used by the participating systems. However, it was not allowed to generate or to make use of large amounts of additional human judgments.

The test sets were annotated manually. For each tuple, seven judges were asked to make a binary decision. Afterwards, the sum of all positive scores yielded a score from 0 to 7. The scores thus necessarily have a certain subjective component. For example, as Amanda Lear grew up in France and Switzerland, but she studied for a while in London, the set contains a score of 2 for the nationality of Amanda Lear being the United Kingdom. For more details: [4].

## 2. OUR APPROACH

Below we describe our approach.

### 2.1 Acquiring Additional Person Files

The organizers provided one large file for all persons, but we wanted to have individual files for each person. We thus crawled Wikipedia[2], where we managed to find the names of more than 99.5% of the given persons. For many persons, there was more than one corresponding article. We crawled and concatenated the content from all corresponding articles following the links on the disambiguation page in order to create a file for a given target person.

Unfortunately, we found out that many articles were deleted from Wikipedia (as not constructive, bad formatted, etc.). Thus, we crawled the DeletionPedia[3], where we found information for about 5% of the persons.

Finally, we crawled DBpedia.[4] We managed to retrieve information for 385,102 out of the 385,426 persons in the training data (99.91%).

---

[1] http://www.wsdm-cup-2017.org/triple-scoring.html
[2] https://www.wikipedia.org/
[3] http://deletionpedia.org
[4] http://wiki.dbpedia.org

## 2.2 Training Data Generation Using Distant Supervision

The organizers provided 162 and 515 training examples for nationalities and professions, respectively. Deliberately this was not enough for proper training, so we created our own training set using distant supervision. Having already downloaded a large number of individual files for persons, we could easily find entirely negative or entirely positive examples.

We took as a *negative* example a tuple (person and entity), where the corresponding entity (profession or nationality) was not mentioned in the person's individual file at all.

We generated a *positive* example tuple (person and entity) when the entity was mentioned in the first sentence of the person's file and no other entities of the same type were mentioned in the file at all. For example, if one person was mentioned as an actor, and no other professions were mentioned in the person's individual file, we considered this to be a positive example.

We generated over 2,000 training examples for both professions and nationalities. We only used the scores 0 and 7, as the above distant supervision approach only yielded entirely positive and entirely negative examples.

## 2.3 Text Normalization

We further used lists of synonyms for professions and adjectives for countries to do text normalization. For example, if a word like *Dutch* occured, we automatically converted it to *Netherlands*. This step was really important as some of our features, e.g., embeddings and occurrence order, relied on exact match.

Moreover, when part of the person's name occurred in the text, we replaced it with the full person name without spaces. This way we can be sure that each occurrence of the person's name is taken into account. This is especially useful for the word embeddings.

## 2.4 Features

We used three types of features.

### 2.4.1 Word2Vec Embeddings

Word2vec is a group of related models that are used to produce word embeddings. These models are shallow, two-layer neural networks, that are trained to reconstruct linguistic contexts of words. Word2vec takes as input a large corpus of text and produces a vector space (typically of several hundred dimensions), with each unique word in the corpus being assigned to a corresponding vector in the space. Word vectors are positioned in the vector space, such that words that share common contexts in the corpus are located in close proximity to one another in the embedding space.

We trained word2vec model on all of the person files, concatenated into one huge document.

### 2.4.2 TF-IDF Feature

In information retrieval, tf-idf, short for term frequency-inverse document frequency reflects how important a word is to a document in a collection or a corpus. It is often used as a weighting factor in information retrieval and text mining [2]. We assign to each term in a document a weight for that term, which depends on the number of occurrences of the term in the document. We would like to compute a score between a query term $t$ and a document $d$, based on the weight of $t$ in $d$.

| | | | |
|---:|---:|---:|---:|
| football | 0.0246 | downs | 0.0057 |
| ball | 0.0142 | leagues | 0.0056 |
| teams | 0.0080 | kick | 0.0051 |
| players | 0.0079 | goal | 0.0049 |
| game | 0.0079 | touchdown | 0.0048 |
| sport | 0.0076 | team | 0.0048 |
| league | 0.0075 | yard | 0.0046 |
| kicks | 0.0069 | line | 0.0046 |
| yards | 0.0064 | scrimmage | 0.0045 |
| defensive | 0.0059 | quarterback | 0.0044 |

**Table 1**: Word weights for *American football player*.

Based on all person files, we determine the top 20 most relevant words for each entity (nationality and profession) and their weights. Let us say that the most common word has weight: 0.023. Then each word in the top 20 list is added to the total result with a number between 0 and 1 = its weight is divided by the max weight for this entity. Table 1 shows word weights for *American football player*.

### 2.4.3 Type-like Occurrence Order Feature

This feature is similar to the one discussed in [3]. It is simpler to implement, but it does not yield worse accuracy. The classification method is the same for professions and nationalities. It consists of the following:

- **Normalization** The most important part of the pre-processing phase, which was performed only for this feature, is that all quoted phrases were removed before the calculation of the corresponding person's nationality/profession score. For example, if a film actor has a role in a film, called "The Mechanic", he is probably not a mechanic. Similar reasoning applies for the case when searching for a person's nationality.

- **Score calculation** The first occurrence of the corresponding entity in the text will get the highest score (which is 7), the next one - 6, until the end of the file or until a score of 0 is reached.

- **Score persistence** In order to avoid the persistence of a large amount of scores, we calculate the score at runtime.

## 2.5 Learning Method

We model the task as a regression problem. In particular, we use linear regression with the above-described features: Word2Vec, TF-IDF, and Type like occurrence order. However, our training examples were only perfect or negative matches (0 and 7).

The type-like occurrence order feature matched this 100% and the linear regression did not give any weight to the other features. We trained linear regression and combined the results equally: final result = 0.5 * type-like occurrence order + 0.5 * linear regression. Table 2 shows the average error on the training data.

## 2.6 Workflow

For our main run, we implemented the following workflow, which is executed for each tuple person–entity:

| Feature | Nationalities err | Professions err |
|---|---|---|
| Word2Vec | 2.63 | 2.11 |
| TF-IDF | 4.06 | 1.93 |
| Occurrence order | 1.41 | 2.00 |
| Linear regression | 2.46 | 1.93 |
| **Combined** | **1.65** | **1.67** |

Table 2: Average error on the training set for each feature type.

1. If we have no file about the person, we give it a score of **3**, i.e., the middle value. If it does exist, go to 2.

2. Check the relation based on the train data generation. If it is certain that the tuple is positive or negative, give it a score of 7 or 0, respectively. Else, go to 3.

3. Use the combined feature. Give the tuple a score equal to 0.5 * occurrence order + 0.5 * Linear regression. I.e., we use equal weights. In future work, we can learn the relative weights as part of the process of training/tuning the model.

4. Round the score from step 3 to the closest integer in the [0;7] interval.

## 3. RESULTS AND ANALYSIS

The organizers used three evaluation metrics:

1. **Accuracy**: the percentage of triples for which the score computed by the system differs from the score in the ground truth by at most 2.

2. **Average score difference (asd)**: for each triple, take the absolute difference of the system score and the score from the ground truth; add up these differences and divide by the number of triples.

3. **Kendall's Tau**: for each relation, for each subject, compute the ranking of all triples with that subject and relation according to the scores computed by the system and the score from the ground truth. Compute the difference of the two rankings using Kendall's Tau.

Our team was given the internal name of *goosefoot*, and we were ranked 9th, 7th and **1st** on the above measures. Table 3 shows the results ordered by Kendall's Tau. Note that some of the teams returned results in [2;5] rather than in [0;7], which significantly improved their accuracy (as this improved their chances of being within no more than 2 points away from the ground truth). We did not do this. Overall, we believe that Kendall's tau is a very appropriate measure for this task, and it is the standard in related tasks such as machine translation evaluation [1].

## 4. CONCLUSION AND FUTURE WORK

We have described the system of the Sofia University's *goosefoot* team for the WSDM Cup 2017 - Triple scoring task. Our approach is based on distant supervision, having individual files for each person and combining simple features in order to achieve meaningful results.

Further development includes experimenting with character-level features and deep learning.

| Team | Accuracy | Avg. score diff. | Tau |
|---|---|---|---|
| **goosefoot** | 0.75 | 1.78 | **0.31** |
| cress | 0.78 | **1.61** | 0.32 |
| bokchoy | **0.87** | 1.63 | 0.33 |
| chaya | 0.70 | 1.81 | 0.34 |
| cabbage | 0.74 | 1.74 | 0.35 |
| chicory | 0.63 | 1.97 | 0.35 |
| kale | 0.69 | 1.85 | 0.36 |
| lettuce | 0.82 | 1.76 | 0.36 |
| gailan | 0.70 | 1.84 | 0.37 |
| chickweed | 0.77 | 1.87 | 0.39 |
| fiddlehead | 0.73 | 1.70 | 0.40 |
| radicchio | 0.80 | 1.69 | 0.40 |
| bologi | 0.68 | 1.91 | 0.41 |
| catsear | 0.80 | 1.86 | 0.41 |
| cauliflower | 0.75 | 1.87 | 0.43 |
| samphire | 0.78 | 1.88 | 0.44 |
| celosia | 0.69 | 1.93 | 0.45 |
| rapini | 0.73 | 2.03 | 0.45 |
| yarrow | 0.60 | 2.04 | 0.45 |
| endive | 0.55 | 2.49 | 0.46 |
| pigweed | 0.74 | 1.94 | 0.48 |

Table 3: Evaluation results.

Our approach also lacks flexibility and generality. It works well when it has access to individual person file. If there is no file for the corresponding person, it cannot predict anything. Another feature could be trained on large amount of trusted independent text. This way, a larger variety of people and professions could be predicted.

## 5. ACKNOWLEDGEMENTS

This research was performed by a team of students from the MSc programs in Computer Science in the Sofia University "St Kliment Ohridski".

We thank our teachers and mentors (Preslav Nakov, Ivan Koychev, Yasen Kiprov) for the support and guidance to our team participation at the WSDM Cup 2017.

This work was partially supported by project ITDGate funded by the Bulgarian NSF under grant number DN 02/11-2016.

## 6. REFERENCES

[1] O. Bojar, Y. Graham, A. Kamran, and M. Stanojević. Results of the WMT16 metrics shared task. In *Proceedings of the First Conference on Machine Translation*, pages 199–231, Berlin, Germany, 2016.

[2] H. S. Christopher D. Manning, Prabhakar Raghavan. *An Introduction to Information Retrieval.* Cambridge University Press, Cambridge, England, 2009.

[3] E. H. Hannah Bast, Björn Buchhold. Relevance scores for triples from type-like relations. *ACM SIGIR Conference on Research and Development in Information*, pages 243–252, 2015.

[4] S. Heindorf, M. Potthast, H. Bast, B. Buchhold, and E. Haussmann. Wsdm cup 2017: Vandalism detection and triple scoring. 2017.